\def\dsum{\mathop{\displaystyle \sum }}

\documentstyle[aps,pre,preprint]{revtex}

\newcommand{\be}{\begin{equation}}
\newcommand{\ee}{\end{equation}}
\newcommand{\ba}{\begin{eqnarray}}
\newcommand{\ea}{\end{eqnarray}}
\newcommand{\DS}{\displaystyle}

\begin{document}
\bibliographystyle{unsrt}

\title{
Geometrical Properties of Cumulant Expansions
} 

\author{K. Kladko and P. Fulde}

\address{Max-Planck-Institute for Physics of Complex Systems, Bayreuther
Str. 40 H.16, D-01187 Dresden, Germany}

\maketitle

\begin{abstract}
Cumulants represent a natural language for expressing macroscopic properties
of a solid. We show that cumulants are subject to a nontrivial geometry. 
This geometry
provides an   intuitive understanding of a number of cumulant relations which had been obtained 
so far 
by using algebraic considerations. We give general expressions for  their infinitesimal 
and finite transformations and represent a cumulant wave operator
through an integration over a path in the Hilbert space. Cases 
are investigated where this
integration can be done exactly. An expression of the ground-state 
wavefunction in terms of the cumulant wave operator is derived. In the second part
of the article we derive  
the cumulant counterpart of Faddeev`s equations and show its connection to
the method of increments.
\end{abstract}

\begin{section}{Introduction}

In statistical physics, quantum mechanics, quantum field theory
and many other fields of  modern theoretical physics 
we constantly face objects of the type
\begin{equation}
F=\ln{\langle e^{\lambda_{1}A_{1}+\lambda_{2}A_{2}+...}\rangle}.
\label{1}
\end{equation}  
Here $A_{1}$, $A_{2}$, ... are abstract elements, $\lambda_{1}$, $\lambda_{2}$, ... 
are numbers, $\langle... \rangle$ means some kind 
of averaging and $e^{(...)}$ denotes a generalized exponential function.

Assuming the average of 1 to be nonzero, we conclude that 
$F$ is an analytic function of the $\lambda_{i}$ in a vicinity of zero. Therefore, it has a Taylor expansion with respect
to the $\lambda_{i}$. This expansion is called {\it cumulant expansion}, the coefficients of the expansion are called {\it cumulants}.

{\it Cumulants (semiinvariants)} have been known long before in mathematical statistics and probability theory.
Kubo \cite{kubo} pioneered in demonstrating  their usefulness in diverse branches of physics.
Considered as abstract objects cumulants form a special kind of noncommuting algebra.
In his original paper Kubo applied relations found  in this algebra 
to  a diverse set of physical problems. 
In particular, he gave  
an elegant derivation of the Ursell-Mayer expansion for classical and quantum gases  
that is usually obtained  by  much longer  diagram considerations, he rederived Goldstone
perturbation theory for the ground-state of a Hamilton operator, and investigated the
motion of an oscillator with a random frequency. 
Kubo stressed the idea that  "the generalized cumulant expansion provides us with a point of view from which
many existent methods in quantum mechanics and statistical physics can be unified" \cite{kubo}.

Cumulants found  a fundamental application in  solid-state physics \cite{fulde1} (and references therein), \cite{fulde3}, \cite{fulde4},
 \cite{fulde5}, \cite{fulde6}
as a  universal language for expressing properties of a solid in a 
{\it size-consistent} way. 
Expressions for the ground-state energy, expectation values of operators, correlation 
functions  and other
quantities characterizing  a solid were derived by applying the cumulant technique. These expressions
remain valid in the strongly correlated case, when the ground-state of the unperturbed
Hamiltonian  is not a single particle one and the usual diagrammatic theory based
on  Wick theorem either breaks down or becomes very complicated. Also, an important feature is that  any  cumulant expression vanishes for statistically
independent processes. Thus the cumulant technique takes into account only relevant,
statistically connected terms, e.g., of perturbation theory. 

In this article we continue exploring the cumulant technique and show that cumulants
possess a nontrivial geometry. This geometry helps to make clear and intuitive
results obtained before by using algebraic considerations. Geometry in an abstract
sense means studying how the properties of objects change under  transformations \cite{dubrovin}.
First, we find a formula for an infinitesimal transformation of a cumulant, and then
integrate it in order to obtain a finite transformation.
 
It comes out that a transformation of a cumulant expression corresponds to a multiplication
by some {\it cumulant wave operator $\Omega$}. Different analytical expressions for $\Omega$
were obtained before through cumulant algebra \cite{fulde1}. We give a general expression
for $\Omega$ through an integration over a path in Hilbert space, and show  cases
when this integration can be done analytically.
Then we investigate the ground-state wavefunction of a Hamilton operator 
and give an expression for it by applying cumulant technique.
For this we first need to add a new element to the cumulant algebra which is the cumulant 
of the number 1.
Next we show how the cumulant expressions found before can be rederived by using
geometrical considerations. This includes orthogonality relations for $\Omega$,
for the ground-state energy, and for expectation values of operators and correlation functions.
Then we give some new applications of the cumulant technique. 

First we show how 
improved short-time propagators for quantum  Monte Carlo methods can be derived
by using cumulant expansions and derive high-order approximations for
coordinate-momentum and coordinate-coordinate propagators.
Then we consider  the foundation of the method of increments and
extend  this method to excited states of a Hamilton operator.
The method of increments has been proven  a powerful and intuitive way of taking into account electron
correlations in  solids 
\cite{stoll}, \cite{fulde2}, \cite{increment1}, \cite{increment2}, \cite{increment3}, \cite{increment4}. 
The existing theoretical justification of the method of increments is
based on Bethe-Goldstone equations. The $n$-th order Bethe-Goldstone equation is equivalent to a variational
calculation with some trial wavefunction (for the explicit construction of this function see \cite{nesbet}).
This makes the Bethe-Goldstone equations approximate by  nature. Therefore, although the
Bethe-Goldstone equations provide a justification for the method of increments, they are not able
to predict what part of the problem is neglected by performing an  $n$-th order increment calculation.
In other words,  we do not have an exact $n$-th order  equation, such that the $n$-th order increment
equation would be an approximation to this equation. Also we should make a remark that the $n$-th order
Bethe-Goldstone equation is restricted to a vacuum taken as the unperturbed state. Therefore, it is not
applicable when we need to take into account additional correlations starting from an already correlated
state.

On the other hand {\it Faddeev`s equations for the 3-body problem} are well known \cite
{faddeev_equations}, \cite{faddeev} and have been successfully
applied in nuclear physics \cite{bethe} and in solid-state theory 
\cite{igarashi1}, \cite{igarashi2}, \cite{igarashi3}, \cite{igarashi4}.
 Faddeev`s equations are {\it exact}, this feature makes
their application mathematically controllable. We show that we are able to generate a hierarchy
of  Faddeev-like cumulant equations. Taking the equation of the $n$-th order in  the leading approximation
we exactly recover the $n$-th order increment expansion. We can see explicitly from 
the $n$-th order cumulant Faddeev`s equation
what part is neglected by  the increment approximation. Therefore, we generally predict the 
validity and the precision of the $n$-th order increment calculation. 
The cumulant Faddeev`s equations enable us 
to formulate an increment method for  excited states of a Hamilton operator.

Finally, we give the conclusions and point out future directions of research.

\end{section}

\begin{section}{Geometrical properties of cumulants}
\begin{subsection} {Definitions and properties}
We start by  defining the cumulant of $N$ operators \cite{fulde1}. 
\newtheorem{definition}{Definition}
\begin{definition}
The cumulant of $N$ operators is defined as
\end{definition}
\begin{equation}
{
\langle A_{1} ... A_{N} \rangle}^{c}=
{
\frac{\partial}{\partial \lambda_{1}} 
...
\frac {\partial}{\partial \lambda_{N}}
}_{\lambda_{1}=...\lambda_{N}=0}
\ln
{\langle x \mid
               e^{\lambda_{1}A_{1}}
             ...
               e^{\lambda_{N}A_{N}} 
\mid y \rangle}.
\label{2}
\end{equation}

Here $x$, $y$ are vectors belonging to the Hilbert space, we suppose  $\langle x \mid y \rangle \neq 0 $.

We denote this cumulant as ${\langle A_{1} ... A_{N} \rangle}^{c}$ or, when we need to give the averaging
vectors $x$ and $y$ explicitly, as ${\langle x \mid A_{1} ... A_{N}\mid y \rangle}^{c}$.

Cumulants posses a number of properties (see \cite{fulde1}). We briefly repeat them below.  
To simplify notations we consider 
the case  of a two-operator cumulant.

1. Linearity:

\[
{\langle A (\alpha B+\beta C)  \rangle}^{c} = 
\alpha {\langle A B  \rangle}^{c} + \beta {\langle A C  \rangle}^{c}
.\]

2. Independence of the norm of the averaging vectors:
\[
{\langle \epsilon x \mid A B\mid y \rangle}^{c}=
{\langle x \mid A B \mid y \rangle}^{c}
\;\;\;
for
\;
\epsilon \neq 0
.\] 

3. Operators $A$ and $B$ are called {\it statistically independent} with respect to $x, y$ when the average of their product
is equal to the product of their averages, i.e., 
\[
\langle x \mid A B \mid y \rangle  = \langle x \mid A \mid y \rangle \langle x \mid
B \mid y \rangle,\; \langle x \mid  y \rangle = 1. 
\]
Then
\[
{\langle x \mid A B \mid y \rangle}^{c} \equiv 0 .
\]

4. Applying Definition 1 we can obtain explicit formulas for cumulants.
For the case $\langle x \mid y \rangle = 1$ they read as
\[
{\langle x \mid A \mid y \rangle}^{c}={\langle x \mid A \mid y \rangle}
\]
\[
{\langle x \mid A B \mid y \rangle}^{c}={\langle x \mid A B \mid y \rangle}
-
{\langle x \mid A \mid y \rangle}
{\langle x \mid B \mid y \rangle} \: ,\: etc.
\]

5. We need to distinguish the cumulant of two operators $A$ and $B$ and the cumulant of an  operator AB 
considered as an entity, when cumulants are evaluated.
The latter
one we denote as $\langle(AB)^{ \bf{\bullet}} \rangle $.

Also we need to distinguish between the number 1  and the unit operator {\bf 1}. For instance $ {\langle 1 \cdot A \rangle}^{c}=
 {\langle A \rangle}^{c}$ and ${\langle {\bf 1}  A \rangle}^{c} \equiv 0 $. 

We define  {\it formal powers} as ${\langle A^{n} \rangle}^{c} = { \langle \underbrace{A \dots A}_{n} \rangle }^{c} $.
Then the {\it formal algebra} of power series must  be considered. 

Let us formulate an important statement.  

\newtheorem{statement}{Statement}

\begin {statement}
Let H be a Hamilton operator. Let $\psi$ be an eigenvector of this operator with  an  eigenvalue E. Then
for any vector $\psi_{0}$, $ \langle \psi_{0} \mid \psi \rangle \neq 0 $  and for any cumulant series A, not including 
c-numbers the following relations hold
\begin{equation}
\label{3}
{\langle \psi_{0} \mid H \mid \psi \rangle}^{c} = E 
\end{equation}
\begin{equation}
\label{4}
{\langle \psi_{0} \mid A H \mid \psi \rangle}^{c} = 0 .
\end{equation}
Proof:
\end{statement}

The proof of (\ref{3}) comes from the definition of a cumulant of one
operator. Equation (\ref{4}) follows from the fact, that $H$ acting on $\mid \psi \rangle$  
leads to a multiplication by a number and therefore the logarithm
in (\ref{2}) factorizes.

The opposite statement is also true:

\begin{statement}

Let 

\[
{\langle \psi_{0} \mid A H \mid \psi \rangle}^{c} = 0
\]
for any $\psi_{0}$, $ \langle \psi_{0} \mid \psi \rangle \neq 0 $ and for any $A$ not including  c-numbers.

Then $\psi$ is an eigenvector of H.

Proof:
\end{statement}

Let $P$ be a projecting operator on the direction of $\psi$.
Without lost of generality we assume $\langle \psi_{0} \mid \psi \rangle  = 1 $. 
Let us consider the following cumulant: 

\[
{\langle \psi_{0} \mid PH \mid \psi \rangle}^{c}=\langle \psi_{0} \mid PH \mid \psi \rangle -
\langle \psi_{0} \mid P \mid \psi \rangle \langle \psi_{0} \mid H \mid \psi \rangle=
\langle \psi_{0} \mid PH-H \mid \psi \rangle .
\]
Suppose $\psi$ is not an eigenstate of $H$. Then $(PH-H) \mid \psi \rangle \neq 0$. Therefore
we can choose  a vector $\psi_{0}$ such that ${\langle \psi_{0} \mid PH \mid \psi \rangle}^{c} \neq 0$.
But this contradicts the condition of the theorem. Therefore, $\psi$ is an eigenstate
of $H$.
\end{subsection}
\begin{subsection}{Transformations of averaging vectors}
\newtheorem{construction}{Construction}
Here we analyse the basic geometric properties of cumulants. We show that a transformation of averaging  
vectors corresponds to a multiplication by the cumulant wave operator
$\Omega$ (for a definition of $\Omega$ in terms of power series see \cite{fulde1}).

First we consider an { \it infinitesimal} transformation of one of the averaging vectors.
\begin{statement}

Let 

\[ \psi_{B}=e^{\delta S} \psi_{B'} =(1+\delta S)\psi_{B'},
\; \; \delta S=\epsilon S, \; \; \epsilon << 1.
\]

Then

\begin{equation}
{\langle \psi_{A} \mid X \mid \psi_{B} \rangle }^{c}=
{\langle \psi_{A} \mid X(1+\delta S)\mid \psi_{B'} \rangle}^{c}.
\label{5}
\end{equation}

Here X is any formal cumulant series.

Proof:
\end{statement}

The proof follows from the definition of a cumulant. We present it for the case  that  $X$ is a one-operator cumulant.

\[
{
\langle \psi_{A} \mid X \mid  \psi_{B} \rangle 
}^{c}
=
{
\frac{\partial}{\partial \lambda}
}_{\lambda=0}
\ln{
\langle \psi_{A} \mid 
e^{\lambda X} \mid
e^{\epsilon S} \psi_{B'} \rangle
}
=
\]

\[
=
{
\frac{\partial}{\partial \lambda}
}_{\lambda=0}
\ln{
\langle \psi_{A} \mid
e^{\lambda X} \mid
\psi_{B'} \rangle
}
+
\epsilon
{
\frac{\partial}{\partial \epsilon}
}_{\epsilon =0}
{
\frac{\partial}{\partial \lambda}
}_{\lambda=0}
\ln{
\langle \psi_{A} \mid
e^{\lambda X} e^{\epsilon S} \mid
\psi_{B'} \rangle
}
=
\]
\[
=
{\langle \psi_{A} \mid X(1+\delta S)\mid \psi_{B'} \rangle}^{c}.
\]

Here we considered an infinitesimal transformation of the left averaging vector.
In an analogous way, if  for the right averaging vector
\[
\psi_{A}=e^{\delta S} \psi_{A'} =(1+\delta S)\psi_{A'},
\delta S=\epsilon S, \: \epsilon << 1
\] 

then $
{\langle \psi_{A} \mid X \mid \psi_{B} \rangle}^{c}=
{\langle \psi_{A'} \mid (1+{\delta S}^{\dagger} )X\mid \psi_{B} \rangle}^{c}.
$

We conclude that an infinitesimal transformation of the averaging vector result in a multiplication
of the cumulant operator by  $1+ \delta S$. In the following we call this operator 
{ \it infinitesimal cumulant wave operator}.

The next step is going over to a finite transformation by integrating an infinitesimal one.

We consider the following construction:

\begin{construction}
Let $\psi_{A}$, $\psi_{B}$, $\psi_{C}$ be vectors of Hilbert space. Furthermore, let us consider a continuous path in  Hilbert space between vectors $\psi_{B}$ and
$\psi_{C}$. Let us choose $N+1$ points $\psi_{0}$, $\psi_{1}$, ... $\psi_{N}$ on this path such that $\psi_{0}=\psi_{C}$ and $\psi_{N}=
\psi_{B}$ and $N$ operators $\delta S_{i}$ such that $\psi_{i}=e^{\delta S_{i}}\psi_{i-1}$.
\end{construction}

By assuming that the  operators  $\delta S_{i}$  are small and using (\ref{5}) we obtain 

\[
{\langle \psi_{A} \mid X \mid  \psi_{B} \rangle}^{c} \approx
\langle \psi_{A} \mid X \prod_{i=1}^{N}(1+\delta S_{i})\mid \psi_{C} {\rangle}^{c}.
\]

Taking formally the limit $N \rightarrow \infty $ we obtain

\[
{\langle \psi_{A} \mid X \mid \psi_{B} \rangle}^{c} = 
\langle \psi_{A} \mid X \Omega \mid \psi_{C} {\rangle}^{c}
\]

\begin{equation}
\Omega =  \lim_{N \rightarrow \infty} \prod_{i=1}^{N}(1+\delta S_{i}) . 
\label{6}
\end{equation}
Thus we conclude that a transformation of an averaging vector in a
cumulant expression corresponds to a 
multiplication with the operator $\Omega$. In the following we call this
operator  {\it cumulant wave operator}. 

We note that there is a considerable freedom in the Construction 1 
for choosing  the path between vectors $\psi_{B}$, $\psi_{C}$ and therefore the 
operators $\delta S_{i}$.     
 Thus the operator $\Omega$ is not 
defined uniquely. Two different representations for $\Omega$ 
differ by a cumulant operator with a zero contribution to any 
cumulant  expression (such operators can be obtained by integrating 
over a closed path in  Hilbert space and subtracting the number 1).

We also note that $\Omega$ always contains the number 1 as the first term
of the formal series. 

The formula for the finite transformation of the left averaging vector 
is obtained in analogy to (\ref{6}). In that case we have

\begin{equation}
\label{7}
{\langle \psi_{A} \mid X \mid \psi_{B} \rangle}^{c} = 
{\langle \psi_{C} \mid 
{\Omega} ^{\dagger} 
X \mid \psi_{B} \rangle}^{c}.
\end{equation}
\end{subsection}

\begin{subsection}{The cumulant of the number 1}
To complete the definition of the algebra of cumulant series we need to define  the cumulant of the number 1.
By formally putting the number of lambdas in the
Definition 1 to zero we have 
\begin{equation}
\label{8}
{\langle x \mid 1 \mid y \rangle}^{c}=
\ln{\langle x \mid 
 y \rangle} .
\end{equation}

The usefulness of this definition comes from the point that the new object 
we just introduced satisfies the general expression for an infinitesimal
transformation of a cumulant. It can be directly checked that

\[
{
{\langle \psi_{A} \mid 1 \mid  \psi_{B} \rangle }^{c}=
{\langle \psi_{A} \mid 1 (1+\delta S)\mid \psi_{B'} \rangle}^{c}.
}
\]
Therefore when doing transformations we  need not to discriminate between the cumulant of the number 1 
and  "ordinary" cumulants.
\end{subsection}
\begin{subsection}{Superoperators}
In  parts of this article we use cumulants of superoperators,
so we will repeat some considerations from \cite{fulde1}.
{\it A superoperator} is an operator acting in the space of all  
operators, i.e. Liouville space. The cumulant of a superoperator $D$ is
defined in a way completely analogous to the ordinary cumulant.
For instance:
\begin{equation}
\label{9}
{
\langle x \mid  D \mid y  \rangle}^{c}=
{
\frac{\partial}{\partial \lambda} 
}_{\lambda=0}
\ln
{\langle x \mid
               e^{\lambda D}
{\bf 1}\mid y \rangle}.
\end{equation}
Here {\bf 1} is the unit operator of the Liouville space.

We see that the only difference is that we need to 
include the unit operator when defining cumulants.

For some operator $H$ we introduce the superoperator $L$:
 $L=[H,...]$.

Cumulants containing $L$ have the following property  (\cite{fischer})

\begin{statement} 
For any cumulant expression G and for any operator A it is
\begin{equation}
\label{10} 
{ \langle G L A \rangle}^{c}={\langle  {G (L A)}^{\bullet} \rangle }^{c} .
\end{equation}
Proof:
\end{statement}

The existence of $G$ does not play any role in the
following proof and so we omit $G$ 
to simplify the writing.
\ba
{\langle L A  \rangle}^{c} 
&=&
\left. \frac{\partial^2}{\partial\lambda_1 \lambda_2} \right|
_{\lambda_{1}=\lambda_{2}=0}
\ln
{\langle x \mid
             e^{\lambda_{1}L}
               e^{\lambda_{2}A} 
\mid y \rangle}			\nonumber	\\
&=&
\frac{\partial}{\partial \lambda_{1}}
\frac {\partial}{\partial \lambda_{2}}
_{\lambda_{1}=\lambda_{2}=0}
\ln
{\langle x \mid
               (1+\lambda_{1}L)
               (1+ \lambda_{2}A)
\mid y \rangle}			\nonumber	\\
&=&
\frac{\partial}{\partial \lambda_{1}}
\frac {\partial}{\partial \lambda_{2}}
_{\lambda_{1}=\lambda_{2}=0}
\ln
{\langle x \mid
               (1+\lambda_{1}\lambda_{2}LA)
\mid y \rangle}
=
\frac {\partial}{\partial \lambda}
_{\lambda=0}
\ln
{\langle x \mid
               (1+\lambda L A)
\mid y \rangle}
\label{11}
\ea
Here we wrote $\lambda=\lambda_{1} \lambda_{2}$.
For the second cumulant we have
\begin{equation}
{\langle (L A)^{\cdot}  \rangle}^{c}=
\frac{\partial}{\partial \lambda}
_{\lambda=0}
\ln
{\langle x \mid
               e^{\lambda L A}
\mid y \rangle}=
\frac{\partial}{\partial \lambda}
_{\lambda=0}
\ln
{\langle x \mid
               (1+\lambda L A)
\mid y \rangle}.
\label{12} 
\end{equation} 
So the left and right sides of (\ref{10}) are the same.

The superoperator $L$ has the following  property: if the right averaging vector
in a cumulant is an eigenvector of $H$ than $L$ is equivalent to $H$ in the
cumulant expression.
Indeed, $L$ implies a commutation with $H$,
and $AH$ gives zero, because
the right averaging vector is an eigenvector of $H$.
\end{subsection}
\begin{subsection}{Finite Temperatures}
The above results can be extended to  finite temperatures.
Here the averaging implies taking a trace with some other operator
(averaging operator).
Therefore, we define the cumulant of a product of $N$ operators as:

\begin{equation}
\label{13}
{
\langle A_{1} ... A_{N} \mid G \rangle}^{c}=
{
\frac{\partial}{\partial \lambda_{1}} 
...
\frac {\partial}{\partial \lambda_{N}}
}_{\lambda_{1}=...\lambda_{N}=0}
\ln
{\mbox{Tr}\left(
               e^{\lambda_{1}A_{1}}
             ...
               e^{\lambda_{N}A_{N}} 
G\right)}.
\end{equation}
Here $G$ is an averaging operator.
Considering infinitesimal transformations of $G$ of the type  $G= e^{\delta S} G' $ we arrive at 
formulas which are identical to those of the zero temperature case.

Let us derive the cumulant expression for the free energy of the system.

Let be $H=H_{0}+H_{1}$ where $H_{0}$ is an unperturbed Hamiltonian, and  $H_{1}$ is a perturbation.

Then 
\ba\label{14}  
F &=& -\beta \ln Tr (e^{- \beta H}) 	
   = -\beta{ \langle 1 \mid e^{- \beta H} \rangle }^{c}	\nonumber  \\
  &=& -\beta { \langle 1 \mid e^{- \beta (L+H_{1})}e^{- \beta H_{0}} \rangle 
		}^{c}
   =  -\beta { \langle e^{- \beta (L+H_{1})} \mid e^{- \beta H_{0}} \rangle 
	}^{c}. 
\ea
In the first step we have used the definition of a cumulant of the number 1 and then 
transformed the averaging operator from  $e^{- \beta (L+H_{1})}e^{- \beta H_{0}}$
to $e^{- \beta H_{0}}$.

\noindent Then ${\DS 
\delta F = F- F_{0} = -\beta { \langle e^{- \beta (L+H_{1})}-1 \mid e^{- \beta H_{0}} \rangle }^{c}}$.

\noindent This expression  is analogous to the closed loop expansion of the 
free energy (thermodynamic potential) in field theory.

\end{subsection}
\end{section}

\begin{section}{Applications of cumulant geometry}
Here we use the results obtained in the previous section  to describe   
properties of a quantum system. Most of these results were obtained
previously by using algebraic considerations~\cite{fulde1,fulde3,fulde4,fulde5,fulde6}. 
Here we rederive them from
a geometrical point of view.

\begin{subsection}{Energy eigenvalues, orthogonality relations and
expectation values}

\begin {statement}
Let $H$ be a Hamiltonian operator. Furthermore, let $\psi$ be an eigenvector of this operator
 with  an eigenvalue $E$. Let 
$\psi_{0}$ be an arbitrary vector of  Hilbert space   $\langle \psi_{0} \mid \psi \rangle \neq 0$, 
and $\Omega$ a cumulant wave operator transforming  $\psi$ to $\psi_{0}$.
Then for any formal series A, not including 
c-numbers, the following equations hold
\be
\label{15}
{\langle \Psi_{0} \mid H \Omega \mid \Psi_{0}  \rangle}^{c} = E 
\end{equation}
\be
\label{16}
{\langle \Psi_{0} \mid A H \Omega \mid \Psi_{0}  \rangle}^{c} = 0. 
\end{equation}
Proof:
\end{statement}

By taking equations (\ref{3}), (\ref{4}) and transforming the right 
averaging vector from $\Psi$ to $\Psi_{0}$ we immediately  
obtain (\ref{15}), (\ref{16}).

Using Statement 2 and transforming  $\Psi$ into $\Psi_{0}$ we find that
the opposite statement is also true, i.e., if $\Omega$ transforms 
$\Psi$ into $\Psi_{0}$ and 
(\ref{16}) holds for any A then $\Psi$ is an eigenvector of H.

Next we consider  expectation values of different physical quantities.

\begin{statement}
Let $\Psi$ be a vector, with $\langle \Psi \mid \Psi \rangle = 1$, let 
$\Psi_{0}$ be another vector and let $\Omega$ be a wave operator transforming $\Psi$ into
$\Psi_{0}$. Then

\be
\label{17}
\langle \Psi \mid G \mid \Psi \rangle ={ \langle \Psi_{0} \mid \Omega^{\dagger} G \Omega \mid \Psi_{0} 
\rangle
}^{c}.
\end{equation}
Proof:
\end{statement}

First we note that $  \langle \Psi \mid G \mid \Psi \rangle ={ \langle \Psi \mid G \mid \Psi \rangle}^{c}$. 
Transforming the right and left averaging vectors from $\Psi$ to  
$\Psi_{0}$ we arrive at (\ref{17}).

\end{subsection}
\begin{subsection}{Explicit forms of the $\Omega$ operator}
We consider  cases when the integration (\ref{6})  can be done
analytically.

\begin{statement}

Let $\Psi_{B}=e^{S} \Psi_{C}$ where $S$ is an operator of Liouville space. 

Then one of the possible forms for the wave operator
transforming $\Psi_{B}$ into $\Psi_{C}$ is 
\be
\label{18}
\Omega = e^{S}.  
\end{equation}
In the last expression $e^{S}$ should be understood in terms of a series expansion.

Proof:
\end{statement}

We take $\delta S_{i}$ in (\ref{6}) as:
\[
\delta S_{i}= S/N.
\]
Then we have for $\Omega$
\[
\Omega =  \lim_{N \rightarrow \infty} \prod_{i=1}^{N}(1+ S/N)=
e^{S}.
\]
We have used  a  representation of the formal exponential series 
known from algebra.
It is possible to prove a more general statement:
\begin{statement}

Let $\Psi_{B}=e^{g(S)} \Psi_{C}$ where $S$ is an operator of Liouville space, and g(S) is an analytic function
with
g(0)=0.

Then one of the possible forms for a wave operator 
transforming $\Psi_{B}$ to $\Psi_{C}$ is
\begin{equation}
\label{19}
\Omega = e^{g(S)}.
\end{equation}
In the last expression $e^{g(S)}$ should be understood  in terms of a series expansion. 

Proof:
\end{statement}

We use for  $\delta S_{i}$ in (\ref{6}) the expansion:
\[
\delta S_{i}= g'(iS/N)S/N
\]
Then we can write 
\[
\Omega= \lim_{N \rightarrow \infty} \prod_{i=1}^{N} 
\left[ 1+ \frac{S}{N} g^\prime\left( \frac{iS}{N}\right) \right].
\]
It is known from  the theory of  series expansions (see for instance \cite{lang})
that this expression tends to
\[
\Omega=\exp\left[ g(S) \right] \quad . 
\]
The question arises under which conditions  an analytic function $f(S)$ can be represented by $e^{g(S)}$.
The theorem for Weierstrass products \cite{lang} states that a necessary and sufficient condition is that
function $f(s)$ does not have zeroes in the complex plane.

For instance, the function $1+S$ has a zero in the complex plane and therefore cannot be represented 
by $e^{g(S)}$. Therefore, we cannot state that if $\Psi_{B}=1+S \Psi_{C}$ then $\Omega=1+S$.

Let us now rederive an operator $\Omega$ transforming the unperturbed
ground-state $\mid \psi_{0}\rangle$ of the Hamiltonian $H_{0}+V$  
to the true ground-state $\mid \psi \rangle$ derived in \cite{fulde1}.
We assume the nondegenerate case.

First we make a remark that 
\be
\lim_{t \rightarrow \infty} e^{-Ht}
\mid \psi_{0} \rangle = e^{-Et} \langle \psi \mid \psi_{0} \rangle     
\mid \psi \rangle.
\ee
Here $E$ is the ground-state energy of $H$. We assume that $\mid \psi \rangle $ and $\mid \psi_{0} \rangle $ have
 nonzero overlap.
Because cumulants do not depend on the norm of the averaging
vectors  the prefactor in front of $\mid \psi \rangle$ is not
important and we can write
\be
\lim_{t \rightarrow \infty} e^{-Ht}\mid \psi_{0} \rangle =\mid \psi \rangle.
\ee

Then using (\ref{18}) we write

\be
\Omega=\lim_{t \rightarrow \infty}\mid e^{-Ht}).
\ee

Taking the   Laplace transform with respect to the variable $z$  and multiplying by $z$ to get the
constant part of the expression we obtain

\be
\Omega= \lim_{z \rightarrow 0}\mid \frac {z} {z-H})=
\lim_{z \rightarrow 0}\mid 1+ \frac {1} {z-H} H)=  
\lim_{z \rightarrow 0}\mid 1+ \frac {1} {z-H} V) .
\ee
This expression coincides with the one\cite{fulde1}.
In the last line we took into account the fact that $\mid \psi_{0}
\rangle$
is an eigenvector of $H_{0}$.

\end{subsection}
\begin{subsection}{Wavefunction} 
Here we consider the question how to obtain the wavefunction from the
$\Omega$ operator.

\begin{statement}
Let $\Psi$ and $\Psi_{0}$ be vectors of Hilbert space, and $\Omega$  
be a cumulant wave operator transforming $\Psi$ to $\Psi_{0}$. Then
there exists an expression for $\Psi$ through $\Psi_{0}$ and $\Omega$ that
is  be derived below.

Proof:
\end{statement}

We use the fact that 
\[
\nabla_{\DS \Psi_{0}} {\langle \Psi_{0} \mid 1 \mid \Psi \rangle}^{c}=
\nabla_{\DS \Psi_{0}} \ln \langle \Psi_{0}\mid \Psi \rangle =
\frac { \mid \Psi  \rangle}{\langle \Psi_{0} \mid \Psi \rangle}=
\mid \Psi_{norm} \rangle. 
\]
Here 1 is number 1 and $\mid \Psi_{norm} \rangle $ means that $\mid \Psi \rangle $ is normalized by the condition
$ \langle \Psi_{0} \mid \Psi \rangle =1 $.

By transforming the left  side  from $\Psi$ to $\Psi_{0}$ we
have:

\be
\label{3-7}
\Psi_{norm}=\nabla_{left  \;  \Psi_{0}} {\langle \Psi_{0} 
\mid \Omega \mid \Psi_{0} \rangle}^{c}.
\end{equation}

Here $left \: \Psi_{0}$ means that we should take the gradient only with respect to the
left $ \Psi_{0} $.

\end{subsection}

\begin{subsection}{Green Functions}

In order to  treat Green functions we need to generalize
the definition of cumulants by including cases  with 
the averaging operators inside.

We explain this extended definition by giving an example:

\be
{\langle u\underline{A}w{\underline B}  \rangle}_{c}=
{
{\frac{\partial}{\partial \lambda_{1}} \frac {\partial}{\partial \lambda_{2
}}}}_{\lambda_{1}=\lambda_{2}=0}
\ln{\langle x \mid u e^{\lambda_{1}A} w
e^{ \lambda_{2}B}
 \mid y \rangle} .
\ee

This is the cumulant of operators $A$ and $B$ with averaging
vectors $\mid x \rangle $ and $\mid y \rangle $  outside and averaging
operators $u$ and $w$ inside.

The usefulness of this definition is seen by noticing, that
we  can write a Matsubara Green`s  function \cite{abrikosov} the following
way (we assume $\langle \psi \mid \psi \rangle =1) $:

\begin{eqnarray}
R_{ij}(\tau)= {\langle \psi \mid c_{i} e^{\tau H} c^{+}_{j} e^{- \tau H} \mid \psi \rangle
}=
{\langle \psi \mid c_{i} e^{\tau H} c^{+}_{j} e^{- \tau H} \mid \psi \rangle}-
{\langle \psi \mid c_{i}  \mid \psi \rangle}
{\langle \psi \mid  c^+_j \mid \psi \rangle}= \nonumber \\ 
{\langle \psi \mid { c_{i}} e^{\tau H} {c^{+}_{j}} e^{
- \tau H} \mid \psi \rangle}^{c} 
=  
{\langle \psi \mid {c_{i}} e^{\tau L} {c^{+}_{j}} 
\mid \psi \rangle}^{c}.
\end{eqnarray}

Therefore, we can write a Matsubara Green function as a 2-operator cumulant.
We can 
find formulas for transformations of averaging operators which are
similar to the formulas for transformations of averaging vectors.

As a result we obtain the  following expression for the Laplace transform
of $R_{ij}$

\be
R_{ij}(z)= (\Omega \mid ( \frac {1} {z-L} c_{i})^{\bullet} c_{j}^{+} \mid \Omega).
\ee

This result coincides with the one obtained in \cite{fulde6,fulde1}  
by resummation of a power
series.

In a similar way expressions for higher-order Green functions can be obtained.
\end{subsection}

\begin{subsection}{Short-time propagators}

The Diffusion Quantum Monte Carlo technique uses Feynman`s path integration
for the evaluation of physical properties of a system. An 
important issue is the quality of a {\it short-time propagator}
by means of which we obtain the evolution of the system after a small
period of time.
M. Suzuki \cite{Suzuki} first proposed to calculate improved
short-time propagators using cumulant expansions. Below we     
explicitly calculate improved coordinate-momentum and coordinate-coordinate
propagators for a general physical system with a Hamilton operator
 $H=p^2/2+V(q)$.
The {\it real-space propagator}  is defined as a matrix element of
the evolution operator between two $\delta$-functions representing
the eigenstates of the position operator \cite{schulman}. Because two delta 
functions situated at different points of the real space have  zero
overlap, we are not able to apply cumulants directly to 
the real space. The idea is then to extend the method  used by
Faddeev \cite{faddeev_lectures} to obtain the first-order propagator. First we 
calculate
the phase-space propagator, and then by taking  
the Fourier transform come back to the real space. 

The {\it phase-space propagator}  is defined as \cite{faddeev_lectures}
\be 
\label{21}
F(p,q,t)=\langle p \mid  e^{-iHt/\hbar} \mid q \rangle =
\exp \left[{\langle p \mid e^{-iHt/\hbar} \mid q \rangle}^{c} \right] .
\end{equation}

Consider $H=p^2/2+V(q)$. The above expression allows us
to evaluate an improved short-time propagator up to
any power of $t$. Let us calculate it to order 
$t^2$. Direct calculation gives:
\be
\label{22}
F(p,q,t)=(2 \pi)^{-1/2} \exp \left[-i p q \hbar^{-1}    - i (p^2/2+V(q)) t \hbar^{-1}-\frac{1}{4}( 
2 p \hbar  V^{\prime}(q)-V^{\prime \prime}(q))t^2 \right].
\ee
Going over to the (q',q'') propagator we have:

\be
\label{23}
F(q^\prime, q^{\prime\prime},t) = \langle q^{\prime\prime} \mid e^{ iHt} 
\mid q^\prime \rangle =
 {(2 \pi)}^{-1/2} \int  e^{ipq''} F(p,q')dp \quad .
\ee

By inserting (\ref{22})  and evaluating the integral over $p$ we obtain the
short-time real space propagator:

\be
\label{24}
F(q',q'',t) = \frac{1}{  \sqrt{(2 \pi i \hbar  T)}}
\exp\left[ 
i \frac{
\left[ q''- q' + (1/2) t^2 V'(q') \right]^2}{2t \hbar } -
i t V(q') \hbar^{-1}  -1/4 t^2 V''(q')
\right].
\ee

In  a similar manner the next order short-time propagators can be computed.

Let us discuss the physical meaning of the expression (\ref{24}). We have two additional
terms in the propagator as compared to the usual leading order  expression 
derived in \cite{feynman}.
The term $(1/2) t^2 V'(q')$  corresponds to the fact that the velocity 
of the particle following a Feynman path is not constant, changing 
due to the force $V'(q')$ and therefore leading to a correction of 
the kinetic energy term. The meaning of the term $1/4 t^2 V''(q')$
will be considered later.

Let us use the above results to obtain a high-temperature expansion for 
the partition function of a quantum system.

The partition function is defined as \cite{feynman}:

\begin{equation}
Z= Tr \exp(-\beta H)= \int \langle q \mid \exp(- \beta H ) \mid q \rangle dq .  
\end{equation}

For small $\beta$ we can use the expansion (\ref{24})  setting 
$t=-i\beta$. Then we have to  order $\beta^2$:

\begin{equation}
Z= \frac{1}{  \sqrt{(2 \pi  \hbar^2 \beta )}}
\int \exp\left[
-
\beta  V(q)  -(1/4) \hbar^2 \beta ^2  V''(q) \label{tunneling} 
\right] dq .
\end{equation} 

To the first order in $\beta$ we obtain  a classical expression for a partition
function, this means that for high temperatures the average energy of the system is
large and we can neglect quantisation effects.

Let us discuss the meaning of the  correction $ 1/4 \hbar^2 \beta ^2  V''(q)$.
Suppose the potential $V(q)$ has a very high but very narrow peak. Then the region of the peak will
be prohibited in  a classical theory because $V(q)$ is large.
On the other hand, 
for  the quantum case there will be some probability for the system to be in the region
of the peak due to  tunnelling. For a high and narrow peak $V''(q)$ will be large and negative.
Now looking at the expression (\ref{tunneling}) we see that $1/4 \hbar^2 \beta ^2  V''(q)$
has then a sign opposite to $\beta  V(q)$ so the potential is effectively smeared out.
Therefore, we see that $ 1/4 \hbar^2 \beta ^2  V''(q)$ term describes an effective smearing
of the potential due to the quantum tunnelling.
  
The expansion for a partition function obtained above is a high temperature expansion  
(in powers of $\beta$). We note that this expansion does not coincide with the quasiclassical
expansion (in powers of $\hbar$). Although both types of expansions give the classical
formula as the leading term, the higher-order  terms are different. The expansion in powers of $\hbar$ 
was first obtained by  Wigner \cite{wigner}. To the second order in $\hbar$ it can be written as:

\begin{equation}
Z= \frac{1}{  \sqrt{(2 \pi  \hbar^2 \beta )}}
\int \exp\left[
-
\beta  V(q)  -(1/12) \hbar^2 \beta ^2  V''(q) +(1/24) \hbar^2 \beta^3 V'(q)^2
\right] dq .
\end{equation}
 
The formulas for improved short-time propagators can be useful 
to  improve the speed and the precision of  numerical Quantum Monte Carlo calculations \cite{Suzuki}.
They  could also improve analytical calculations that make use of  
short-time propagators. 

We  remark that the above technique applies only
when the short-time propagator has a well-defined Fourier transform
with respect to the coordinate. This is usually satisfied for potentials
approaching zero or a constant at infinite distance. This condition includes the majority
of potentials used for Monte Carlo calculations of quantum gases and
liquids. 
The case when the propagator cannot be Fourier transformed with
respect to the coordinate is subject to future considerations.

\end{subsection}
\end{section}
\begin{section}{Method of Increments}
\begin{subsection}{Derivation of the  Method of Increments starting from 
Faddeev`s Equations 
}
It happens frequently that we deal with a Hamiltonian consisting
of different parts and that these parts considered separately can be solved
analytically or numerically. 

The first example of such a situation
is the{\it 3-body problem}. Three particles interact with each other by
two-particle potentials. Considering each pair of particles
separately we face a 2-body problem which can be solved analytically.
Then the question is how we can use the information given by 
these solutions to construct an approximation to the solution
of the whole 3-body problem.

The second example are  quantum chemistry calculations where we 
need to find an approximate solution for a large molecule
or a cluster having available numerical calculations for its subparts, i.e.,
chemical bonds.

For the 3-body problem L. D. Faddeev  invented {\it coupled integral equations}
that connect the unknown 3-body scattering matrix to 2-body scattering
matrices that can be found analytically \cite{faddeev_equations}, \cite{faddeev}.

In quantum chemistry the { \it method of increments} is well known, 
that expresses the ground-state energy in terms of a sum of 
local increments given by sets of 1, 2, 3 etc. chemical bonds.

Below we show that Faddeev`s equations and the method of increments
are closely related. The increment method can be obtained as
a leading order approximation to {\it cumulant Faddeev-like equations}
that we derive  below.

We start by writing the  wave operator in the form \cite{fulde1}, 

\be
\label{25}
\left| \Omega \right) =\lim _{z\rightarrow 0}\left| 1+\frac 1{z-H}V\right) .  
\ee
We also define a  scattering operator $S$, which is small for small $V,
$
\be
\label{26}
 S=\Omega -1=\lim _{z\rightarrow 0}\sum_{n=1}^{\infty} \left| \left( \frac 1{z-H_0}V\right)
^n\right) .  
\ee
Let us first consider the case when the perturbation consists of three parts.
\begin{statement}
Consider a Hamiltonian with three perturbations, $H_1$, $H_2$ and $H_3$
\be
\label{27}
H=H_0+H_1+H_2+H_3 . 
\ee
Let $S_1$, $S_2$, $S_3$ be scattering operators for $H_0+H_1$, $H_0+H_2$ and $%
H_0+H_3$ respectively. 

Then the scattering operator $S$ of the Hamiltonian $H$ can be found from the solution
of the formal equations

\be
\begin{array}{c}
\label{28prime}
T_1=S_1\left( 1+T_2+T_3\right)  \\
T_2=S_2\left( 1+T_1+T_3\right)  \\
T_3=S_3\left( 1+T_1+T_2\right)  \\
S=T_1+T_2+T_3 .
\end{array}
\ee
Proof:
\end{statement}

We write $S$ in a power series as

\be
S=\sum\limits_{n=1}^\infty \left( \frac 1{z-H_0}\left( H_1+H_2+H_3\right)
\right) ^n 
\ee
and define

\be
A_1=\frac 1{z-H_0}H_1,\: A_2=\frac 1{z-H_0}H_2, \: A_3=\frac 1{z-H_0}H_3. 
\ee

Then

\be
\begin{array}{c}
S=\sum\limits_{n=1}^\infty \left( A_1+A_2+A_3\right) ^n=\sum A_1\ldots
+\sum A_2\ldots +\sum A_3\ldots = 
T_1+T_2+T_3
\end{array} . \label{sum}
\ee

We divided the terms in the power series into three groups, the terms in the first
group begin with $A_1$, the ones in the second group with $A_2$, in the third group
with $A_3$. We call these three sums $T_1$, $T_2$, $T_3$.
 
We rewrite $T_1$ in the form 

\be
\begin{array}{c}
T_1=\left( A_1+A_1^2+\ldots \right) +\left( A_1+A_1^2+\ldots \right)
A_2+\left( A_1+A_1^2+\ldots \right) A_3+\ldots = \\ 
=\left( A_1+A_1^2+\ldots \right) \left( 1+T_2+T_3\right) . 
\end{array}
\ee

The first bracket defines $S_1$. The expressions for $T_2$, $T_{3\text{ }}$ are
similar. This results in  Faddeev`s equations (\ref{28prime}).

In the original work of Faddeev $S_1$, $S_2$, $S_3$ correspond to  two-body
scattering matrices coming from three pairs of particles in the 3-body problem, and $S$
corresponds to the unknown 3-body scattering matrix.

We easily generalize our equations to the case when the perturbation has $N$
parts, in that case  we have $N$ equations of the type

\be
\label{29}
T_i=S_i\left( 1+\sum_{j\neq i}T_j\right).  
\ee
Assuming $S_i$ to be small, the equations can be solved by iterations. Then 
$S$ is given by (\ref{sum}). In the
leading 
approximation we have

\be
\label{30}
T_i=S_i,\: S=\sum S_i . 
\ee

By using the formulas $E=(H \mid \Omega)$ and  $\Omega = 1 + S$ where $E$ is the ground-state energy of $H$ we can
write 

\begin{eqnarray}
\label{30prime}
\delta E=E-E_0= \sum \delta E_i, \\
\delta E_i=(H|S_i).
\end{eqnarray}

Here $E_0$ is the ground-state energy of $H_0$. 
The equation (\ref{30prime}) gives the simplest  
example of a increment method. We call   
the $\delta E_i$`s   first-order increments.

Let us state the result: we start  from Faddeev`s equations,
solve them in the leading order approximation and obtain a first-order
increment method.

Below we show that we can formulate higher-order Faddeev`s equations.
Taking the N-th order Faddeev`s equation and solving it in the
leading order  approximation we obtain exactly the N-th order increment method.
Therefore Faddeev`s equations and increment methods are closely related.

As an example  we derive the second-order Faddeev`s equations.
We again take a Hamiltonian consisting
of  $H_0$ part and $N$ perturbations

\be
\label{31}
H=H_0+H_1+\ldots +H_N . 
\ee
But now we additionally  assume that we can solve analytically or numerically all the Hamiltonians of  the form $%
H_0+H_i+H_j$, i.e., that we know the scattering operators $S_{ij}$ for these
Hamiltonians.

Then we  write again the expression for $S$,

\be
\begin{array}{c}
\label{32}
S=\sum\limits_{n=1}^\infty \left( A_1+\ldots +A_N\right) ^n= \\
\sum A_1\ldots
A_2\ldots +\sum A_2\ldots A_3\ldots +\sum A_1\ldots A_3\ldots +\ldots
+\sum_{n=1}^\infty A_1^n+\ldots \sum_{n=1}^\infty A_N^n= \\ 
= \sum_{1\leq i,,j\leq N.}T_{ij}+\sum_{n=1}^\infty S_i.
\end{array}
\ee
This resummation is more sophisticated and needs explanations. First, we open
brackets in the sum in order to obtain   a power series with respect to $A_i$'s. For each term
in this power series we look for  the  first  two letters in it, and
regroup terms according to this property. For example, $T_{1,2}$ will be the sum
of all terms in the power series having $A_1$ and $A_2$ as the first two letters, 
so that ,e.g., the term $A_1^3A_2^2A_4$ is included in $T_{1,2}$. 

We are still left with terms containing only one letter like $A_2^4$%
. These terms are not contained in any  of $T_{ij}$'s so we sum them separately, this
explains the result of (\ref{32}).

Now we consider $T_{1,2}$:

\be
\label{33}
\begin{array}{c}
T_{1,2}=\left( A_1A_2+A_1A_2^2+A_1^2A_2+A_1 A_2 A_1 +...\right) \left( 1+\dsum_{\left(
i,j\right) \neq \left( 1,2\right)}T_{ij}+\sum_{i\neq 1,2}S_i\right) = \\ 
= \left( S_{1,2}-S_1-S_2\right) \left( 1+\dsum_{\left( i,j\right) \neq \left(
1,2\right)}T_{ij}+\sum_{i\neq 1,2}S_i\right) . 
\end{array}
\ee
This result is obtained by considerations similar to the previous ones. The first
bracket is a sum of all possible terms consisting of $A_1$`s and $A_2$`s only. The second bracket is
a sum of all possible terms that do not have $A_1$ and $A_2$ as the first two
letters. $S_{1,2}$ denotes the scattering operator for the Hamiltonian $H_0+H_1+H_2$.

The similar equations are obtained for all other $T_{ij}$. Together with 
(\ref{32}) these equations give the second-order cumulant Faddeev`s equations.

Taking again the leading approximation for the equations we have:

\be
\label{34}
S=\sum_iS_i+\sum_{i,j}K_{ij} 
\ee

where 

\be 
\label{35}
K_{ij}=S_{ij}-S_i-S_j .  
\ee
Here $K_{ij}$ is a cumulant operator corresponding to the second-order increment. 

Using the formula for the ground state energy $E=\left( H\mid \Omega \right) $
we arrive at an expression up to the second-order increment expression

\be
\delta E=\sum_i \delta E_i +\sum_{ij} \delta E_{ij} .
\ee

The second order increments $\delta E_{ij}$ are given by

\be
\delta E_{ij}=(H|K_{ij})=(H|S_{ij})-\delta E_{i}- \delta E_{j}
\ee
By continuing this procedure  and taking the leading approximation
to the next contribution we end up with the third-order cumulant expression for $S$

\be
\label{36}
S=\sum_iS_i+\sum_{i,j}K_{i,j}+\sum_{i,j,k}L_{ijk}  
\ee 
\be  
\label{37}
L_{ijk}=S_{ijk}-K_{ij}-K_{ik}-K_{jk}-K_i-K_j-K_k . 
\ee

Again this expression agrees up to  third-order with the energy expression
of the increment method. The same holds true for higher-order contributions.

Next we consider the  error when terms up to the   N-th order
increment are taken into account. From  (\ref{33})  we see that the first bracket
is of second order in the perturbation, while the remaining term  that
we omit in the leading-order  approximation is of  first order, therefore  making 
a leading-order approximation
to the equations finally leads to an error $S$ and $E$ which is of third order. Generally,
inclusion of Nth order increments will lead to an error which is of the order $N+1$. 

Actually, the increment expansion  may converge much more rapidly
and  the perturbation does not need to be small.

Consider for instance a large molecule. The term omitted by the leading order 
approximation to Faddeev`s equations can make a significant change to the result
only when all its parts add coherently; otherwise its contribution will
be close to zero. Consider a term corresponding to a set of K
chemical bonds. If the spatial extent of this set is larger than a 
a characteristic length we called the correlation length $%
\lambda $, then the omitted terms in the Faddeev`s equation will sum
incoherently and can be neglected. Therefore we
can terminate an expansion of the correlation energy in terms of 
increments when the spatial extent of a set of bonds is of the order of the correlation 
length. This is in accordance with the well known idea, that the minimum size to which
we can reduce a system not changing its macroscopic properties is dictated by the
correlation length \cite{wilson}. In order to formulate it more rigorously
we may state that the macroscopic properties of an infinite solid can be described 
by an incremental expansion  with increments taken up to the correlation length.
Therefore incremental methods will work efficiently for systems having a moderate
correlation length, like semiconductors \cite{increment2}.  

\end{subsection}
\begin{subsection}{Method of increments for excited states}
 In the following we formulate a method of increments for excited states.
We begin with   a nondegenerate unperturbed Hamiltonian, so 
perturbations do not split excitation energies, instead they 
only shift them.

Consider again as an example a Hamiltonian with three perturbations
$H=H_{0}+H_{1}+H_{2}+H_{3}$.

Consider an excited state $E_{i}$ of $H_{0}$. 
We want to show that the incremental method for an excited state
can be formulated in a similar way as done before  for the ground-state .

We start from  a formula derived in \cite{fulde4}
for the excited states of a Hamiltonian:

\be
\label{38}
E_i ={ \langle \psi_{i} \mid H \Omega \mid \psi_{i} \rangle}^{c} .
\ee 

Here $ \mid \psi_{i} \rangle $ is an excited state of $H_{0}$ having an overlap
of more than 1/2
with the true excited state we are looking for (for details see
\cite{fulde4}).

Let us apply the incremental expression for $\Omega$ to this formula.

To first order of the  incremental expansion (see (\ref{30})) we obtain 
\be
\label{39}
\delta E_i= \delta E_{i,1}+\delta E_{i,2}+\delta E_{i,3} .
\ee

Here $\delta E_{i,1},\delta E_{i,2},\delta E_{i,3}$ are increments corresponding to the excited state $i$,
\be
\delta E_{i,j} = { \langle \psi_{i} \mid H S_i \mid \psi_{i} \rangle}^{c} .
\ee
In the second order of the incremental expansion we find:
\be
\label{40}
\delta E_i= \delta E_{i,1}+\delta E_{i,2}+\delta E_{i,3}+\delta E_{i,1,2}+\delta E_{i,2,3}+\delta E_{i,1,3}
\ee
where $\delta E_{i,1,2}={ \langle \psi_{i} \mid H S_{1,2} \mid \psi_{i} \rangle}^{c} - \delta E_{i,1} - \delta E_{i,2}$ etc.

In an analogous way next order increment expressions can be obtained.

Let us consider as an example of an increment calculation 
the valence band calculations
done in \cite{graefenstein}.

In a first step self-consistent field CSCF calculations are done and the
energy bands  in SCF approximation are obtained.
Then the SCF states are transformed into  Wannier
states (localized orbitals) and  the residual part of
the interaction $H_{res}=H-H_{SCF}$ is expressed in terms of matrix
elements between the localized bonds.

The increment method up to forth order was used to obtain the valence
band. In this case SCF hamiltonian plays the role of $H_0$ and the matrix elements
of the residual Hamiltonian between pairs of localized bonds play role of perturbations.
In the calculations increments corresponding to NN (nearest neighbours) and NNN (next-nearest
neighbours) were retained and the effect of more distant correlations was approximated by
introducing a polarizable continuum.

The results have been shown to describe with a good accuracy the reduction of the valence bonds  of diamond 
due to correlations  \cite{graefenstein}.

\end{subsection}

\end{section}
\begin{section}{Conclusions}
In this article we considered the question, how geometrical
methods can be used to formulate the theory of cumulant expansions. 
Cumulant geometry simplifies the proofs of many expressions obtained
before by cumulant algebra and makes them transparent. 

Then  we  studied  applications of the cumulant technique to 
short-time propagators and to the method of increments. In both
cases the use of cumulants is a natural choice. A 
hierarchy of cumulant Faddeev-like equations was generated and 
a method of increments for excited states was established.

Future directions of research will include diverse applications of the cumulant
technique to solid state theory. A particularly interesting
question there is how  properties of the infinite solid can
be related to  properties of a finite cluster. This question
is of outmost importance for all numerical calculations.
\end{section}
\begin{section}{Acknoledgments}
We would like to thank Prof. K. Becker, K. Fischer and Tran Mien Tien for stimulating
discussions and Tran Mien Tien for carefully reading and correcting the manuscript.
\end{section}

\end{document}